\documentclass[preprint,aps,showkeys,showpacs]{revtex4-2}   
\usepackage{epsfig}
\usepackage{xcolor}   
\usepackage{amsmath}

\begin{document}

\title{Modulational instability in a quasi-one-dimensional Bose-Einstein condensates}

\author{Sherzod R. Otajonov$^{1,\, 2}$}
\author{Bakhram A. Umarov $^1$}
\author{Fatkhulla Kh. Abdullaev $^{1,\,3}$}


\affiliation{$^1$Physical-Technical Institute of the Uzbek Academy of Sciences,\\
Chingiz Aytmatov Street 2-B, Tashkent, 100084, Uzbekistan}

\affiliation{$^2$ Theoretical Physics Department, National University of \\ Uzbekistan, Tashkent 100174, Uzbekistan}

\affiliation{$^3$ Institute of Theoretical Physics, National University of Uzbekistan, 100174, Tashkent, Uzbekistan}

\begin{abstract}
	In this work, we investigate the modulational instability of plane wave solutions within a modified Gross-Pitaevskii equation framework. The equation features cubic and quartic nonlinearity. It models the behaviour of quasi-one-dimensional Bose-Einstein condensates in symmetric Bose-Bose mixtures of ultra-dilute cold atoms. Our study demonstrates the pivotal role of the competition between mean-field attractions and quantum fluctuation-induced repulsions. This competition significantly affects the emergence and evolution of modulational instability. By employing linear stability analysis, we identify the essential conditions that lead to modulational instability. We find that the stability of plane wave solutions significantly depends on the interaction among system parameters. Further development of the instability leads to the fragmentation of the BEC into a chain of quantum droplets. We calculated the quantity of quantum droplets generated during the nonlinear phase of the instability. Our analytical results are corroborated by numerical simulations of the modified quasi-1D Gross-Pitaevskii equation. These simulations vividly depict the formation, interaction, and coalescence of droplets during the nonlinear phase of modulational instability. The investigation shows that linear stability analysis of the modified Gross-Pitaevskii equation, considering quantum fluctuations, precisely forecasts modulational instability phenomena across different domains of parameter spaces.
\end{abstract}

\maketitle

\section{Introduction}
\label{sec:intro}
The experimental realization of Bose-Einstein condensate (BEC) in dilute atomic gases was a milestone achievement that opened the way for new studies, both experimental and theoretical, in low-temperature condensed matter physics~\cite{Book1, Book2}. In particular, it has been shown that mean field theory, based on the Gross-Pitaevskii equation (GPE), can be successfully applied to explain the existence, stability, and dynamics of BEC~\cite{Book1, Book2}. According to the mean-field theory of BEC, the liquid state of weakly interacting atoms, due to attraction, is not formed in two and three dimensions because of collapse phenomena. Moreover, even the corrections introduced by quantum fluctuations~\cite{LHY}, which lead to an additional weak repulsion term (LHY corrections) in the modified single-component GPE, are not enough to prevent collapse. Recently, a new approach to the problem of stable liquid state formation in two-component BEC was put forward theoretically~\cite{Petrov2015, Petrov2016} and soon implemented experimentally~\cite{Barbut2016, Cabrera2018, Skov2021}. As proposed by Petrov~\cite{Petrov2015}, in symmetric binary BECs, the interspecies and intraspecies interactions can be manipulated in such a way that the final single modified GPE will contain both the attractive mean-field term and the repulsive LHY term. Importantly, the strength of the attractive term can be controlled independently, allowing the formation of dilute liquid droplet states stabilized by quantum fluctuations. The modified three-dimensional (3D) GPE, with a quartic nonlinearity term accounting for quantum fluctuations, was successfully applied to describe the properties of BEC droplets in free space~\cite{Otajonov2022_1}. 

	Additionally, variants of the modified GPE in one dimension and two dimensions, describing quantum droplets in tight-binding external potentials, were derived, as detailed in Ref.~\cite{Luo2021}. Several studies have explored various aspects of low-dimensional BEC, as reported in Refs.~\cite{Kartashov2022, Otajonov2019, Otajonov2022, Li2018, Otajonov2020, Mithun2020, Abdullaev2019}. Specifically, for the two-dimensional (2D) case, reports include modulation instability (MI) and the creation of QDs in binary BEC~\cite{Otajonov2022}, along with investigations into the stationary and dynamic characteristics of QDs and vortices~\cite{Li2018, Otajonov2020} . In one-dimensional (1D) scenarios, studies have detailed the MI and Faraday instability in binary BEC~\cite{Mithun2020, Abdullaev2019}, as well as examined the stationary and dynamic aspects of QDs~\cite{Otajonov2019}.

It was shown that in 1D, the LHY term was modified (introducing quadratic nonlinearity) and changed sign, indicating that the mean-field two-body interaction must be repulsive for quantum droplets to exist \cite{Petrov2016, Otajonov2019, Otajonov2022, Li2018, Otajonov2020, Mithun2020, Abdullaev2019}. In another scenario, if the confinement by the external potential in 1D is loose, then the LHY term remains as it is in the 3D case. Consequently, the modified GPE in quasi-one dimension (quasi-1D GPE)\cite{Debnath2021, Debnath2022} incorporates both cubic mean field and quartic (LHY) terms. 

One method to generate localized waves or solitons is through modulational instability (MI)\cite{Benjamin1967, Ostrovskii1967}, a widespread phenomenon in physics characterized by nonlinear wave equations. Modulational instability enhances small perturbations in a periodic waveform due to nonlinearity, leading to the creation of spectral sidebands and the disintegration of the waveform into a sequence of solitary pulses. The feasibility of MI in a BEC of cold atoms has been theoretically investigated using the mean-field GPE and confirmed experimentally~\cite{Modinst bec1, Modins bec2, Salasnich2003, Abdullaev2009}. More recently, several publications~\cite{Modinst1, Modins2} have explored MI of plane wave solutions in modified GPE with LHY terms. In this work, we focus on MI within the context of quasi-1D GPE featuring quartic nonlinearity. We conduct a linear stability analysis of a periodic plane wave solution analytically and determine the conditions for MI. The analytical results obtained are verified through numerical simulations of the quasi-1D GPE.

\section{The model and results}
\label{sec:model}

	When both components of a binary Bose-Einstein condensate share identical normalization and coupling constants, indicating symmetry, their dynamics in quasi-one-dimensional settings in the presence of quantum fluctuations can be modelled using the following single normalized Gross-Pitaevskii equation, \cite{Debnath2021, Debnath2022, Abdullaev2023}
\begin{eqnarray}
& i \hbar \cfrac{\partial \phi}{\partial T} + \cfrac{\hbar^2}{2 m_0} \cfrac{\partial^2 \phi}{\partial X^2} + \cfrac{2 \hbar^2 \delta a }{m_0 l_{\perp}^2}|\phi|^2 \phi  - \cfrac{512 \sqrt{2} \hbar^2 a^{5/2} }{15 \pi m_0 l_{\perp}^3} |\phi|^3 \phi=0\,,
\label{dimGPE1D}
\end{eqnarray}
where $\phi$ represents the wave function of the condensate, with $T$ indicating time, and $m_0$ denoting the mass of the atoms involved. The term $l_{\perp} = \sqrt{\hbar / m_0 \omega_{\perp}}$ defines the harmonic oscillator length, where $\omega_{\perp}$ refers to the frequency of the transversal trap. The expression $\delta a = -a + a_{12}$ characterizes the residual mean-field interaction effect, distinguishing between the intra-species scattering lengths, denoted by $a_{11} = a_{22} = a$, and the inter-species scattering length, $a_{12}$.

By applying the rescaling transformations $t = T/t_S$, $x = X/x_s$, and $\psi = \phi/\psi_s$, we can express Eq.(\ref{dimGPE1D}) in a dimensionless form:
\begin{equation}
  i \psi_t + \cfrac{1}{2} \psi_{xx} + \gamma |\psi|^2 \psi - \delta |\psi|^3 \psi = 0,
\label{quasi_1D_gpe}
\end{equation}
where the scale parameters are defined as:
$$t_s=\cfrac{2^{16}\, m_0 \gamma^3 a^5}{225\, \pi^2 \hbar \, \delta^2 \delta a^3 }\,, \quad
x_s=\cfrac{256}{15 \pi} \left( \cfrac{\gamma^3 a^5}{\delta^2 \delta a^3} \right)^{1/2}\,,$$  
$$\psi_s=\cfrac{15 \pi \delta a \, l_{\perp} \delta }{256 \sqrt{2}\, a^{5/2} \gamma } \, .$$
The parameters $\gamma$ and $\delta$ may be chosen freely and, within the context of the dimensionless equation, they represent the magnitudes of the residual mean-field interaction and quantum fluctuation terms, respectively. 

Let us look for the homogeneous plane wave solution of  Eq.(\ref{quasi_1D_gpe}) in the following form, indicating an homogeneous distribution of the condensate,
\begin{equation}
  \psi=A \exp(-i \, \mu \, t ) \, .
\label{cw}
\end{equation}
Here $A$ is the amplitude, and $\mu$ the chemical potential. The dependence of the chemical potentials on the amplitudes can be found from Eqs.~(\ref{quasi_1D_gpe}) and (\ref{cw}): 
\begin{eqnarray}
  &\mu=-\gamma A^2+\delta A^3.
\label{omegaj}
\end{eqnarray}

   By applying linear stability analysis, we examine the behavior of small disturbances, $\delta \psi \ll A$, applied to the stationary plane wave solution,
\begin{equation}
  \psi=(A+\delta\psi)\exp(-i \, \mu \, t) \, .
\label{pert}
\end{equation}
The dynamics of small-amplitude $\delta \psi$ are characterized by the equations below:
\begin{eqnarray}
  & i \delta \psi_{t}+ \cfrac{1}{2} \delta\psi_{xx} + (\gamma A^2 - 3 \delta A^3 /2 ) (\delta\psi^*+\delta\psi)=0.
\label{linear}
\end{eqnarray}
The perturbation is expressed as $\delta \psi = u + iv$, leading to the separation of Eq.~(\ref{linear}) into its real and imaginary components. By positing that $(u,v) \sim \exp{(\lambda t+ik x})$, we derive the ensuing characteristic equation:
\begin{eqnarray}
   \lambda^2  + \cfrac{k^2}{2} \left(\cfrac{k^2}{2} -2 \gamma A^2 +3\delta A^3 \right)=0.
\label{lam4}
\end{eqnarray}
Equation~(\ref{lam4}) forms a quadratic equation in terms of $\lambda$, with the solution being
\begin{equation}
  \lambda_{\pm}= \pm \sqrt{ -k^4/4 +k^2 A^2(\gamma -3 \delta A /2 )} .
\label{sol2}
\end{equation}

The growth rate of MI is characterized by the positive real part of the exponents, denoted as $G \equiv \mathrm{Re}(\lambda_\pm) > 0$. For $\gamma,\delta >0$, a plane wave exhibits modulational instability when the subsequent conditions are met: 
\begin{equation}
   |k| \leq k_{cr}, \qquad A < 2 \gamma / 3 \delta,
\label{logarg}
\end{equation}
where the critical value $k_{cr}$, is determined as follows: 
\begin{equation}
  k_{cr} \equiv \sqrt{4 A^2 \gamma -6 A^3 \delta}.
\label{kcr}
\end{equation}

The peak value of the MI growth rate, $G_{max}$, is reached at $k= k_{max}$, which is defined by the following relation:
\begin{equation}
G_{max}={k_{max}^2 \over 2}={k_{cr}^2 \over 4} \, , \qquad  k_{max}={k_{cr} \over \sqrt{2}}\,.
\label{maxkG}
\end{equation}
%

\begin{figure}[htbp]
  \centerline{ \includegraphics[width=6cm]{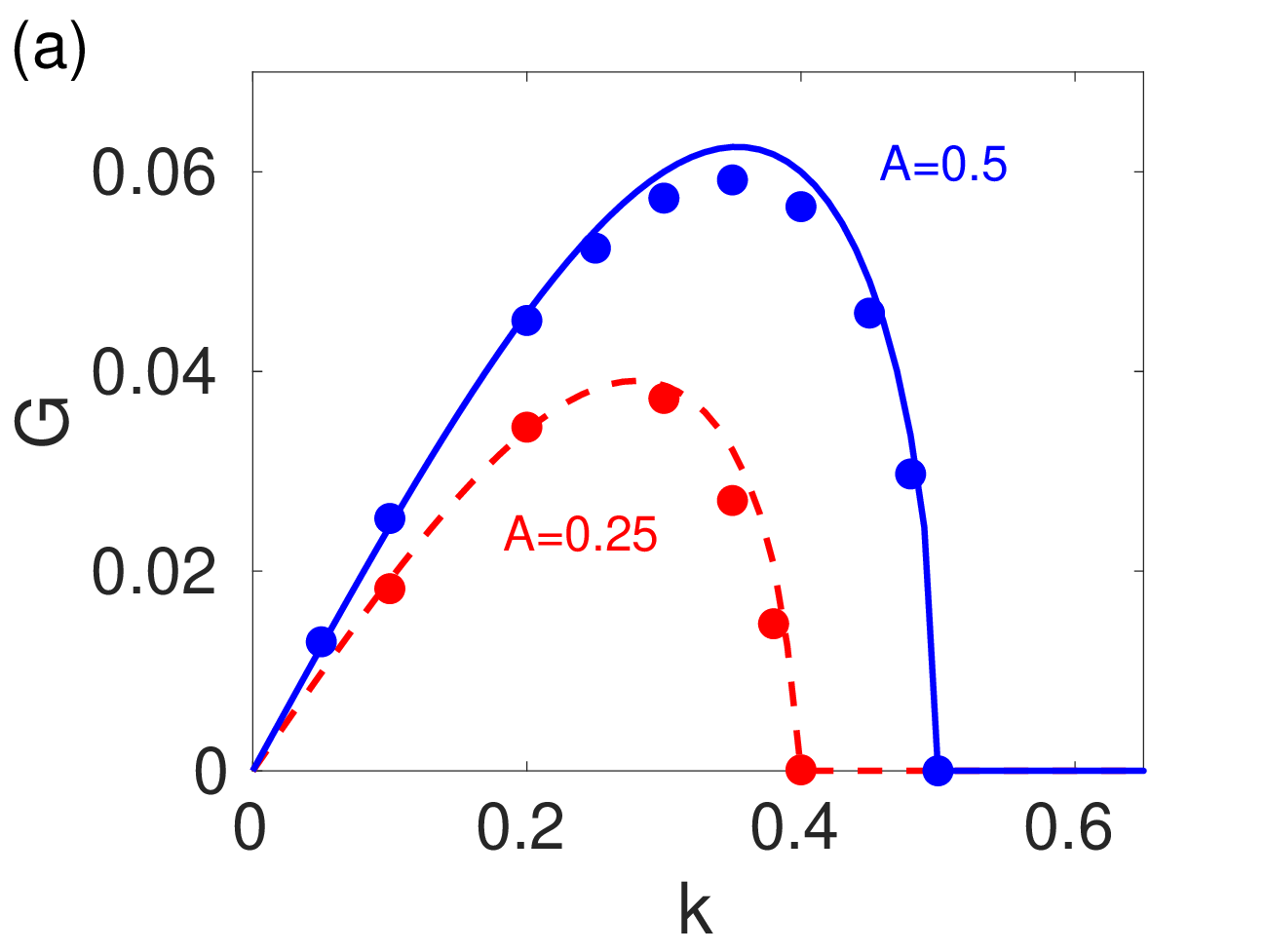}
   \includegraphics[width=6cm]{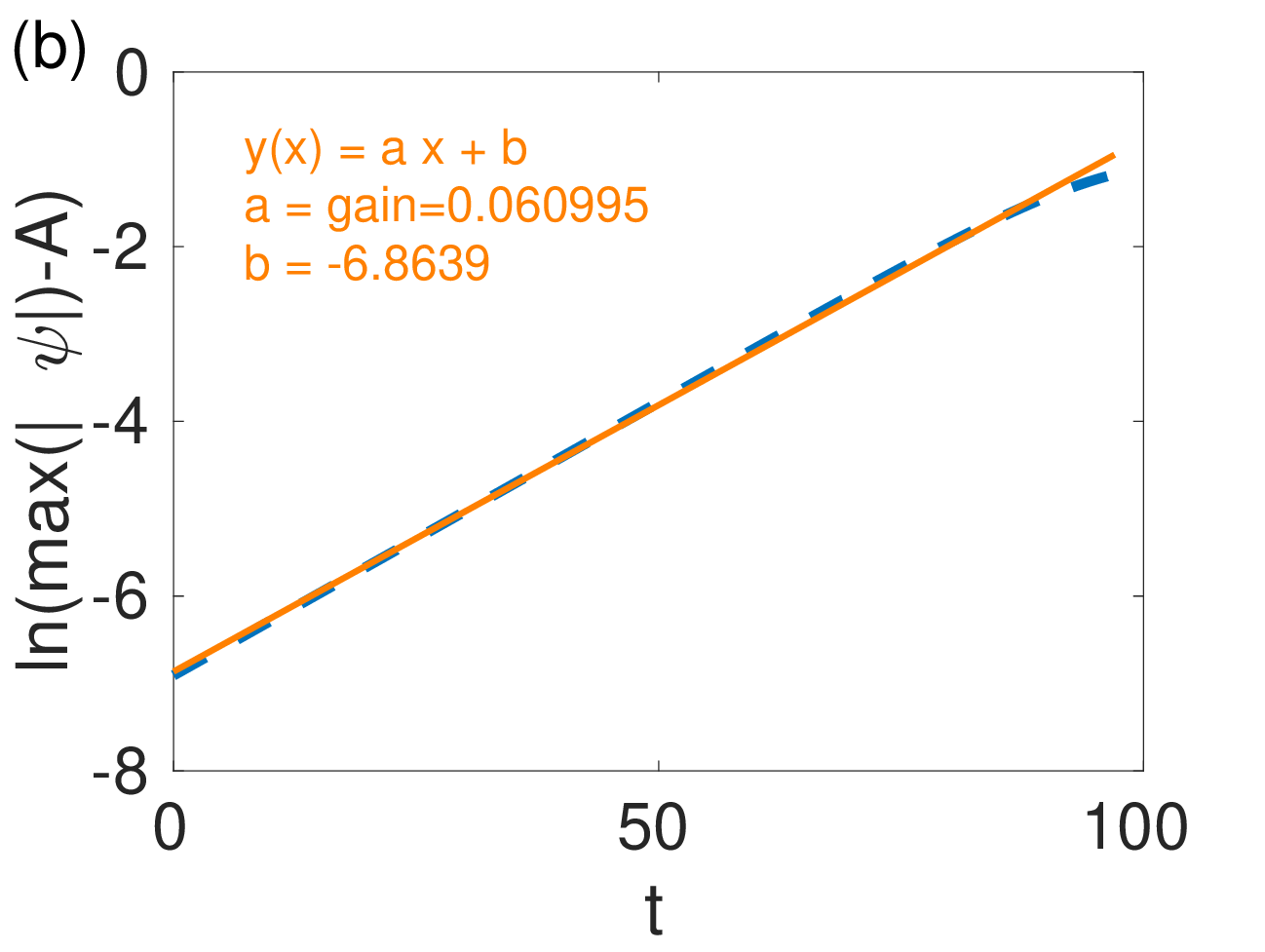} }
\caption{(a) The characteristic MI gain spectrum is illustrated with the upper (blue) line for an amplitude $A=0.5$ and the lower (red) line for $A=0.25$. Points are obtained from numerical simulations based on Eq.~(\ref{quasi_1D_gpe}). (b) The dynamics of the difference in amplitude are plotted on a logarithmic scale for $A=0.5$ and $k=k_{\mathrm{max}}$, where the dashed line represents the outcome of numerical simulations and the solid line indicates the results from linear fitting. The other parameters used are $\gamma=1$ and $\delta=1$.
}
\label{fig-1ab}
\end{figure}
In Fig.~\ref{fig-1ab}(a) the characteristic MI gain spectrums obtained from numerical simulations based on Eq.~(\ref{quasi_1D_gpe}) and from linear stability analysis are shown for two different amplitudes.
Fig.~\ref{fig-1ab}(b) shows the dynamics of the difference in amplitude plotted on a logarithmic scale, where the dashed line represents the outcome of numerical simulations and the solid line indicates the results from linear fitting. The results presented in Figs. ~\ref{fig-1ab}(a,b) confirm that the linear stability analysis describes well the initial stages of modulational instability when amplitude grows exponentially.
  
\begin{figure}[htpb]
  \centerline{ \includegraphics[width=5.5cm]{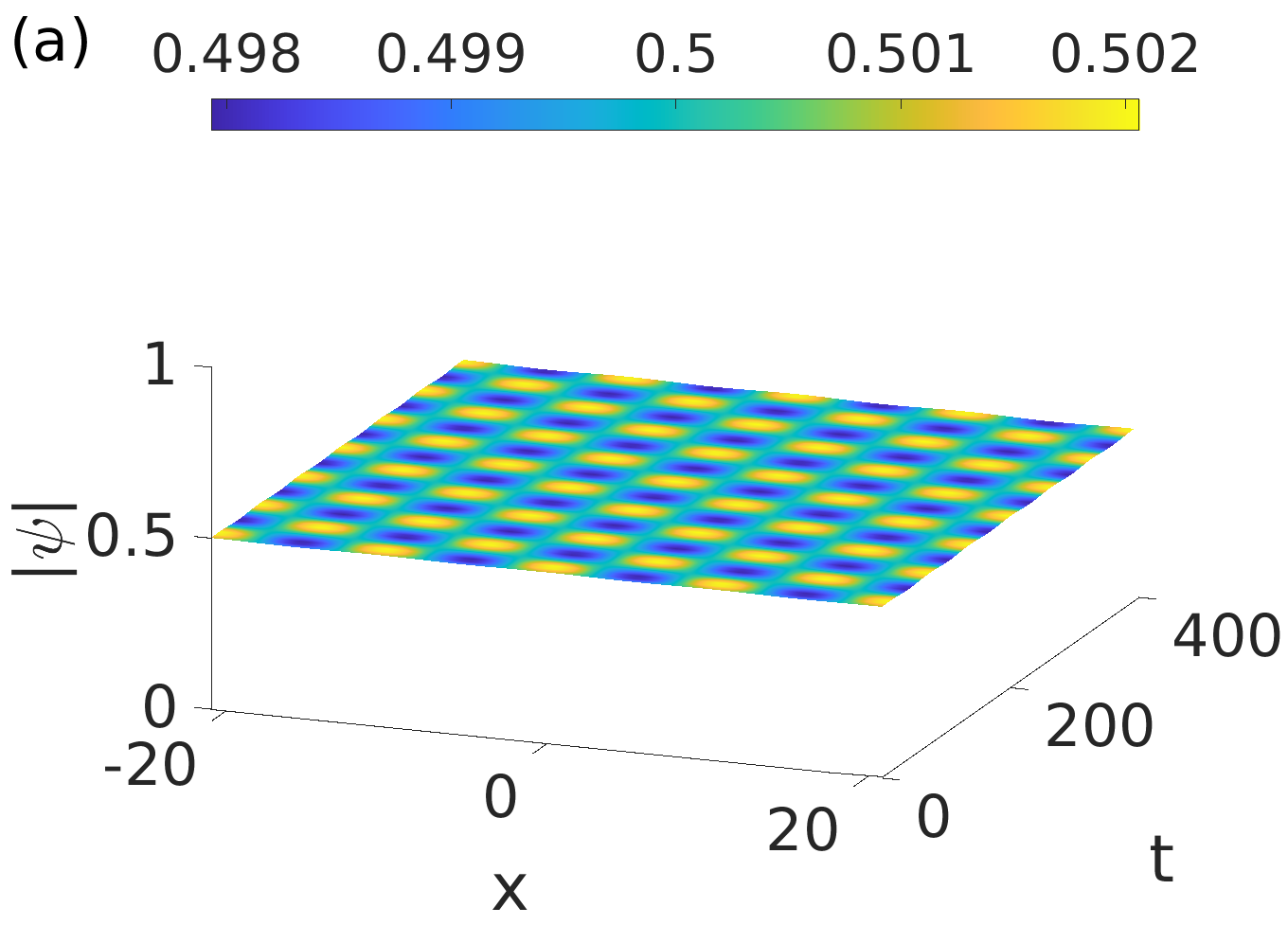}
     \includegraphics[width=5.5cm]{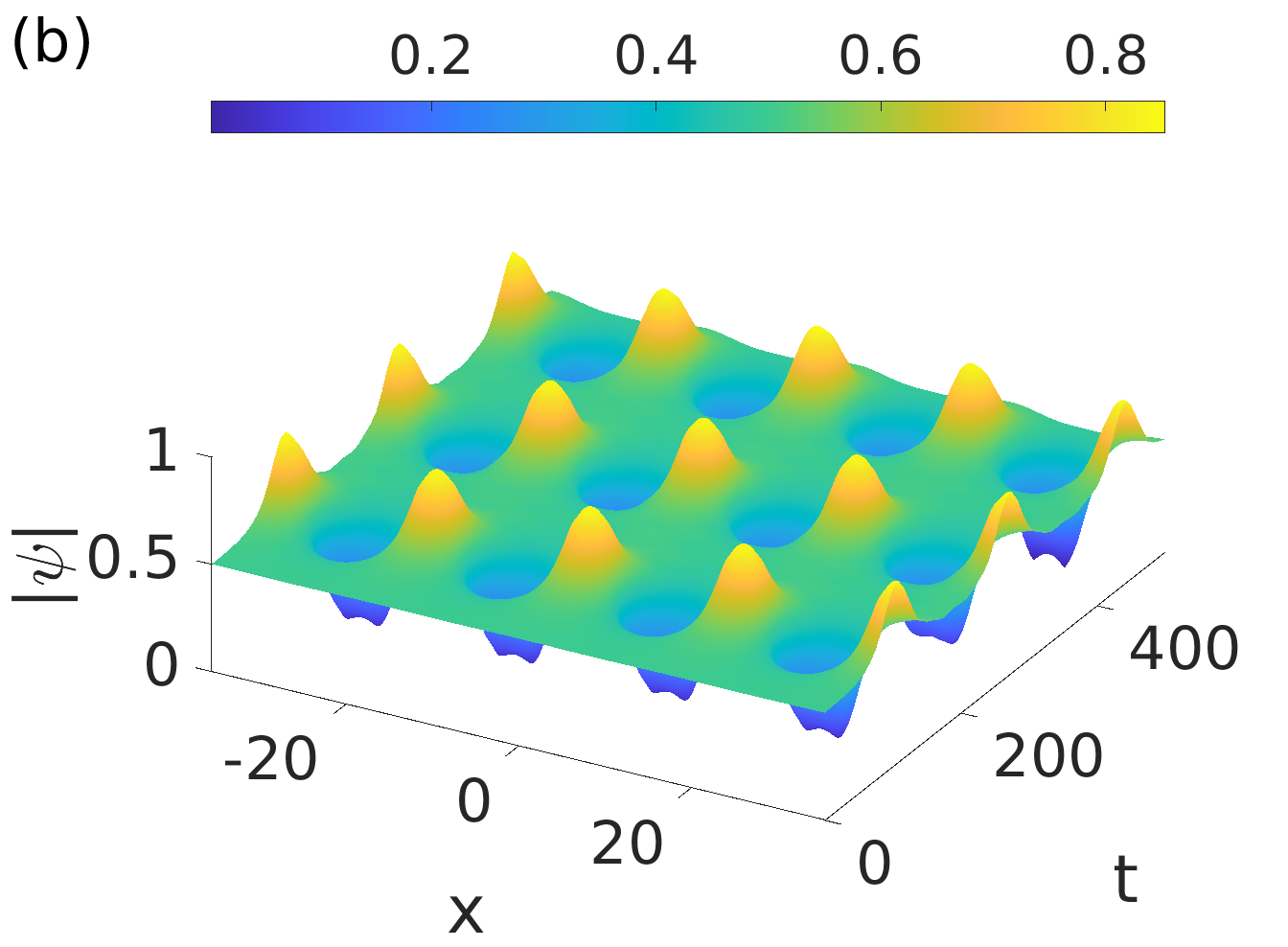}}
\caption{Dynamics of slightly perturbed plane wave solutions. (a) When $k=0.6$, exceeding the critical value $k_{\mathrm{cr}}$. (b) At $k=k_{\mathrm{max}}$, which is less than $k_{\mathrm{cr}}$. The remaining parameters are set as $A=0.5$, $\gamma=1$, and $\delta=1$.}
\label{fig-2ab}
\end{figure}

Also, it is necessary to point out that within the framework of modulational instability gain spectrum analysis presented in Fig.~\ref{fig-1ab}(a), when the wavenumber $k$ exceeds a certain critical point, the MI growth rate diminishes to an inconsequential level, leaving the plane-wave solutions unaltered, as depicted in  Fig.~\ref{fig-2ab}(a). 
%
On the other hand, when $k$ falls below this critical level, the amplitude of a small perturbed plane wave starts to growth exponentially, as illustrated in Fig.~\ref{fig-2ab}(b). It is essential to acknowledge that linear stability analysis merely identifies the preliminary conditions leading to MI and does not elucidate the wave field's further dynamics. The transition to the nonlinear dynamics phase, represented in Fig.~\ref{fig-2ab}(b), is not encompassed within the linear analysis paradigm.

\begin{figure}[htbp]
  \centerline{ \includegraphics[width=5.cm]{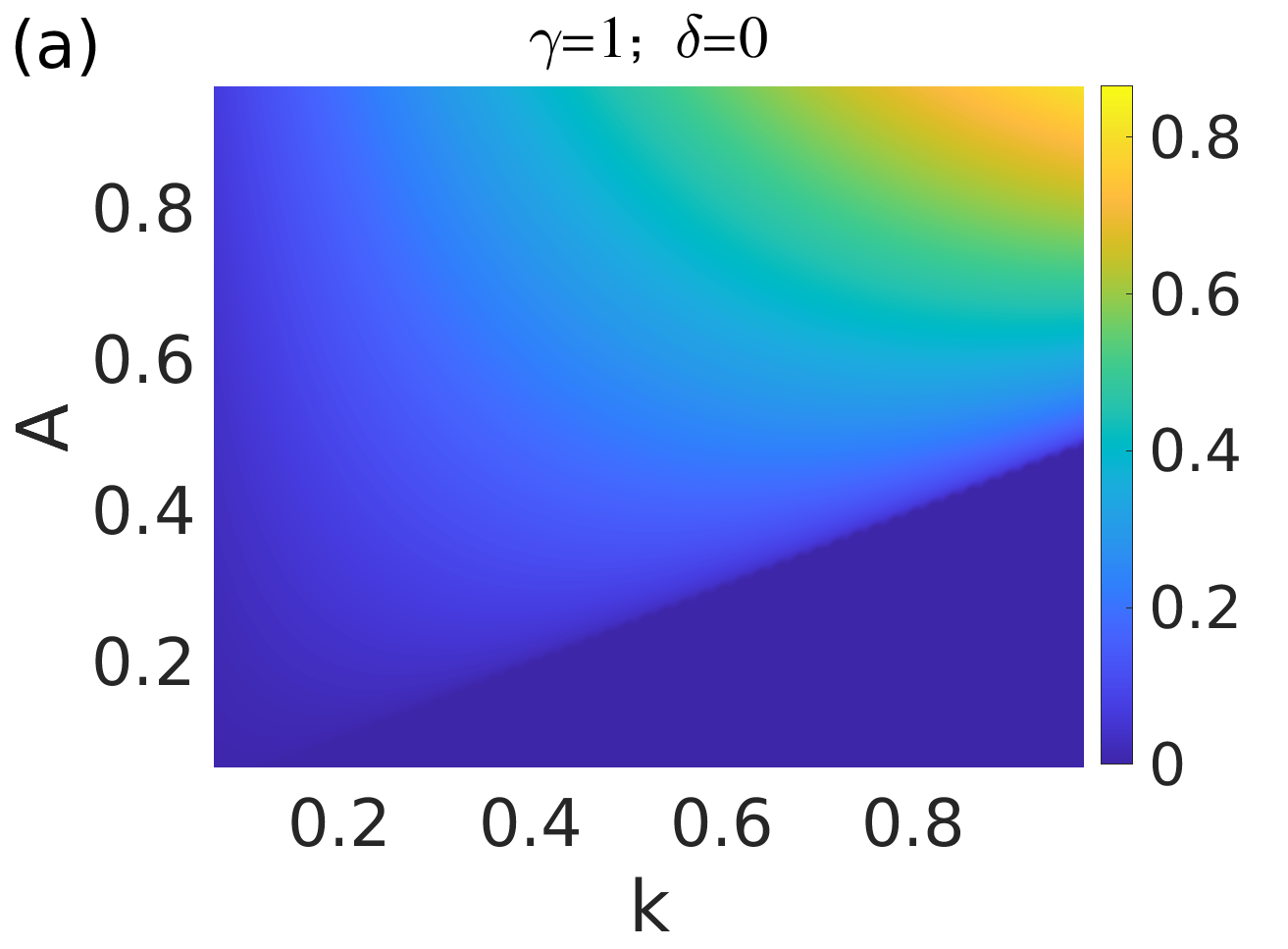}
   \includegraphics[width=5.cm]{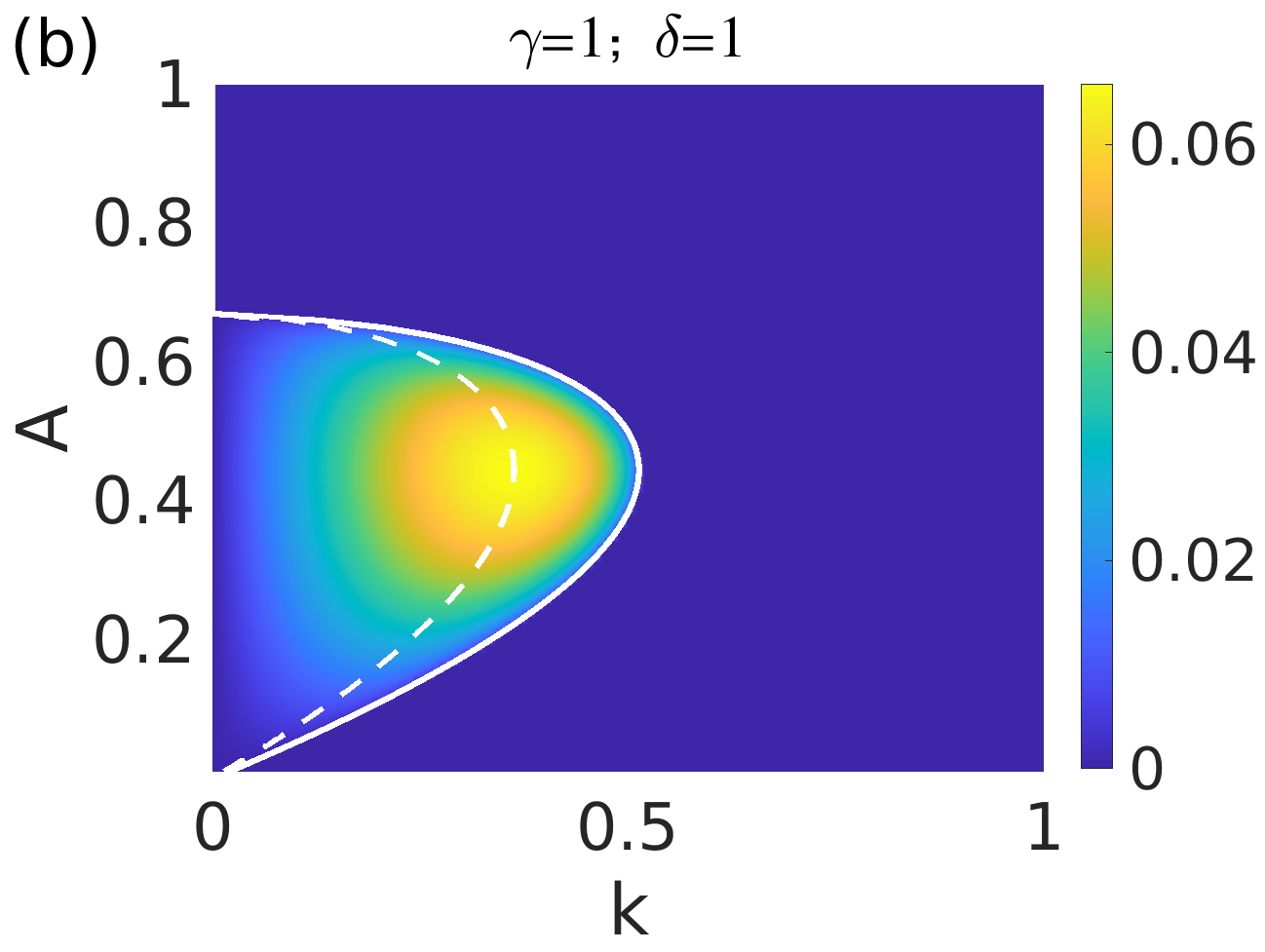} }
\caption{The growth rate $G$ is plotted as a function of $A$ and $k$. (a) corresponds to the scenario where $\gamma=1$ and $\delta=0$. (b) also depicts the case with $\gamma=1$ and $\delta=0$. The solid line shows the MI boundary ($k_{\mathrm{cr}}$), and the dashed line represents $G_{max}$, see Eq.~(\ref{maxkG}).
}
\label{fig-3ab}
\end{figure}

Figure~\ref{fig-3ab}(a) shows the dependence of the MI growth rate on the wave number for ($\gamma, \delta$)=(1,0).  at $\delta=0$ we have the mean field GPE where the domains of MI occurrence are unbounded in the (A,k)-plane. 

From Fig.~\ref{fig-3ab}(b) one may note that when quantum fluctuation is included,  ($\gamma, \delta$)=(1,1),  MI happens only in certain finite intervals.

\begin{figure}[htbp]
 \centerline{ \includegraphics[width=5cm]{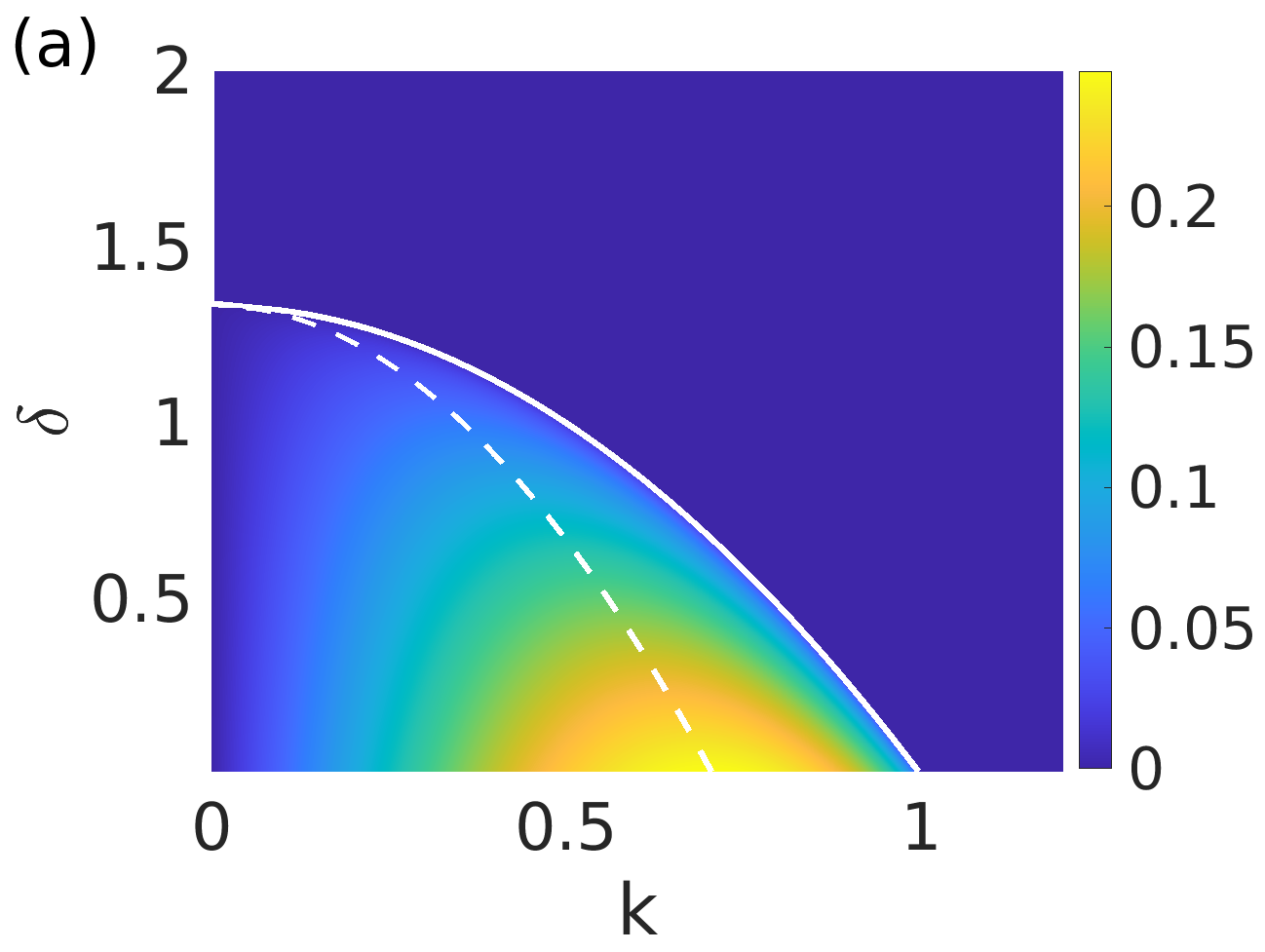}
 \includegraphics[width=5cm]{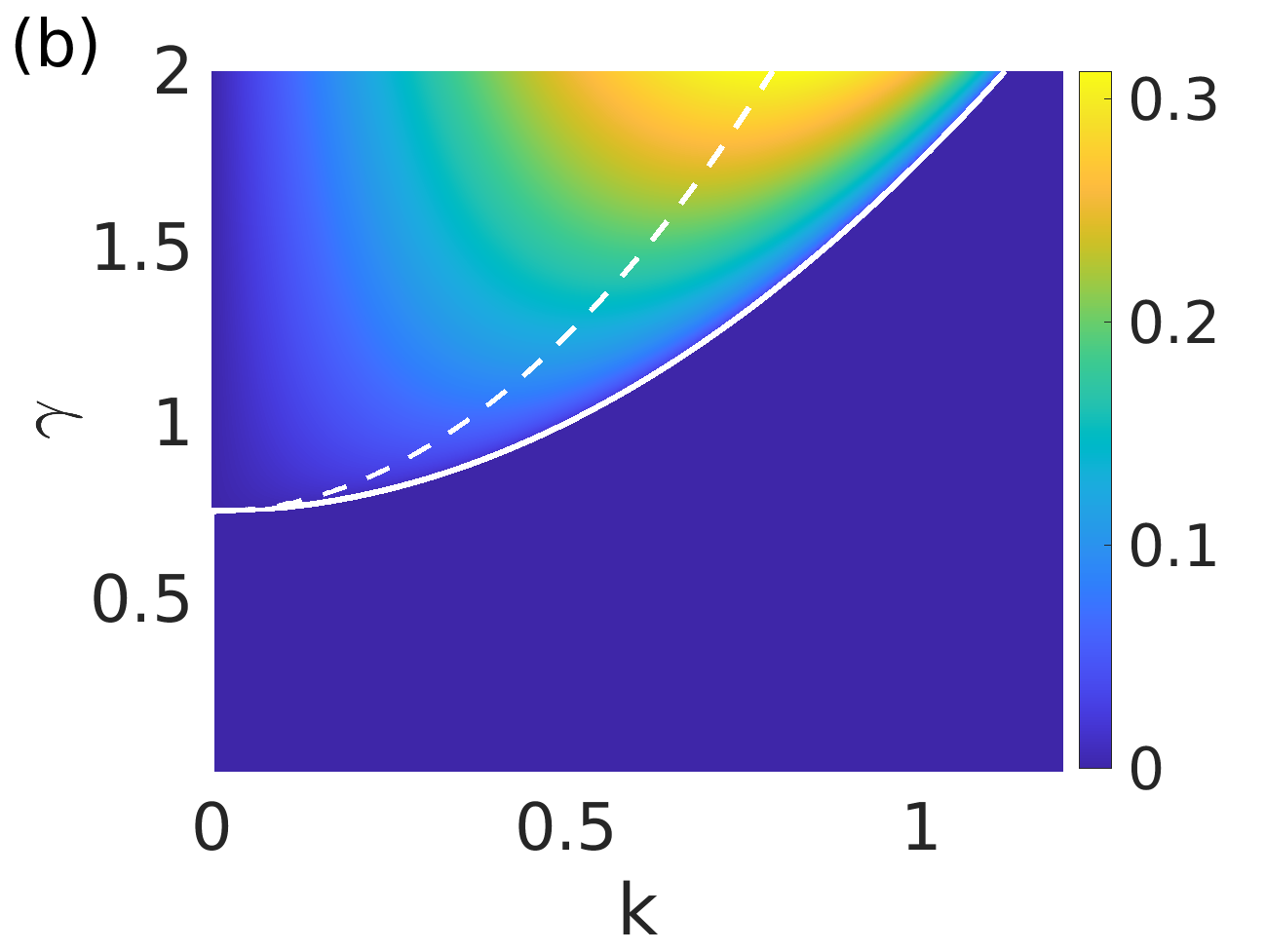} }
\caption{The MI growth rate is represented for $A=0.5$. (a) illustrates this on the $(\delta, k)$-plane with $\gamma=1$, while (b) shows it on the $(\gamma,k)$-plane with $\delta=1$. Both solid and dashed lines correspond to Eq.~(\ref{maxkG}).
}
\label{fig-4ab}
\end{figure}

Figure~\ref{fig-4ab}(a) illustrates the MI growth rate, $G$, as a function of the nonlinearity coefficient $\delta$ and wave number $k$, with a fixed amplitude $A=0.5$ and interaction strength $\gamma=1$. The graph demonstrates that MI occurs within a specific, bounded domain on the ($\delta, k $)-plane. Both the solid and dashed lines likely represent the analytical prediction of MI onset, illustrating a threshold beyond which MI is observable. The bounded domain suggests that MI is only feasible under certain conditions of $\delta$ and $k$, highlighting the intricate balance required for the emergence of MI in this system.

Figure~\ref{fig-4ab}(b) shows the MI growth rate, $G$, on the ($\gamma, k$)-plane, with $\delta$ held constant at 1 and amplitude $A=0.5$. Unlike Figure 4(a), the areas where MI occurs are not restricted, indicating that variations in the interaction strength $\gamma$ and wave number $k$ offer a wider range of parameters for observing MI. The continuous and dashed lines, adhering to Eq.~(\ref{maxkG}), mark the analytical boundaries for MI, suggesting an unbounded domain of MI occurrence. This implies that adjusting $\gamma$ provides flexibility in achieving MI across a broader parameter space.

	Figure 5(a) depicts the MI growth rate, $G$, within the ($\delta, \gamma$)-plane for a fixed amplitude $A=0.5$. The colour bar represents the maximal values of the growth rate, indicating regions of both stability and instability. Areas of high growth rate correspond to conditions favourable for MI, while transitions between colours illustrate the delicate thresholds separating stable and unstable regimes, underscoring the nuanced dynamics controlled by quantum fluctuations and mean-field effects.
	
\begin{figure}[htbp]
\centerline{ \includegraphics[width=5cm]{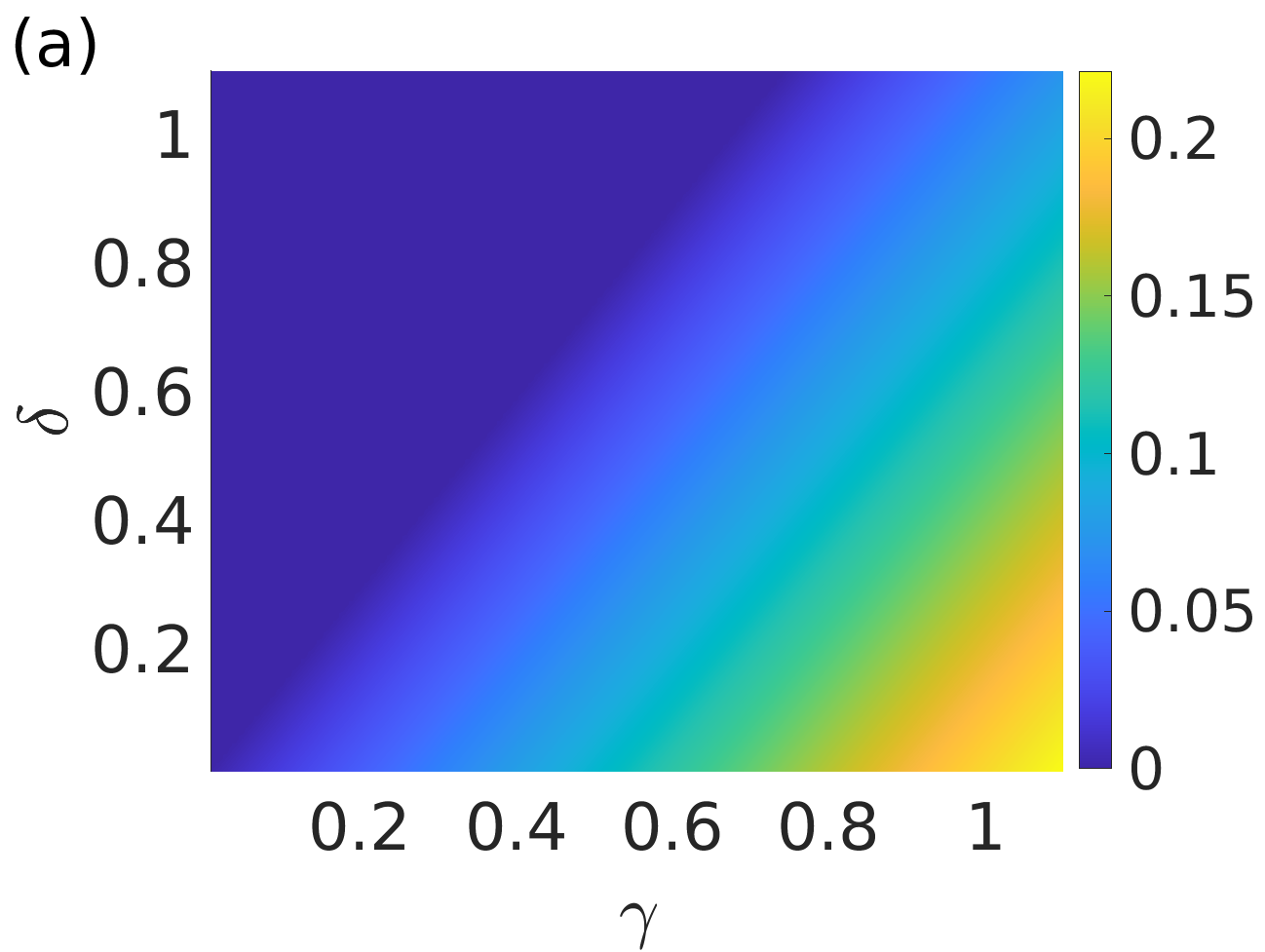}
\includegraphics[width=5cm]{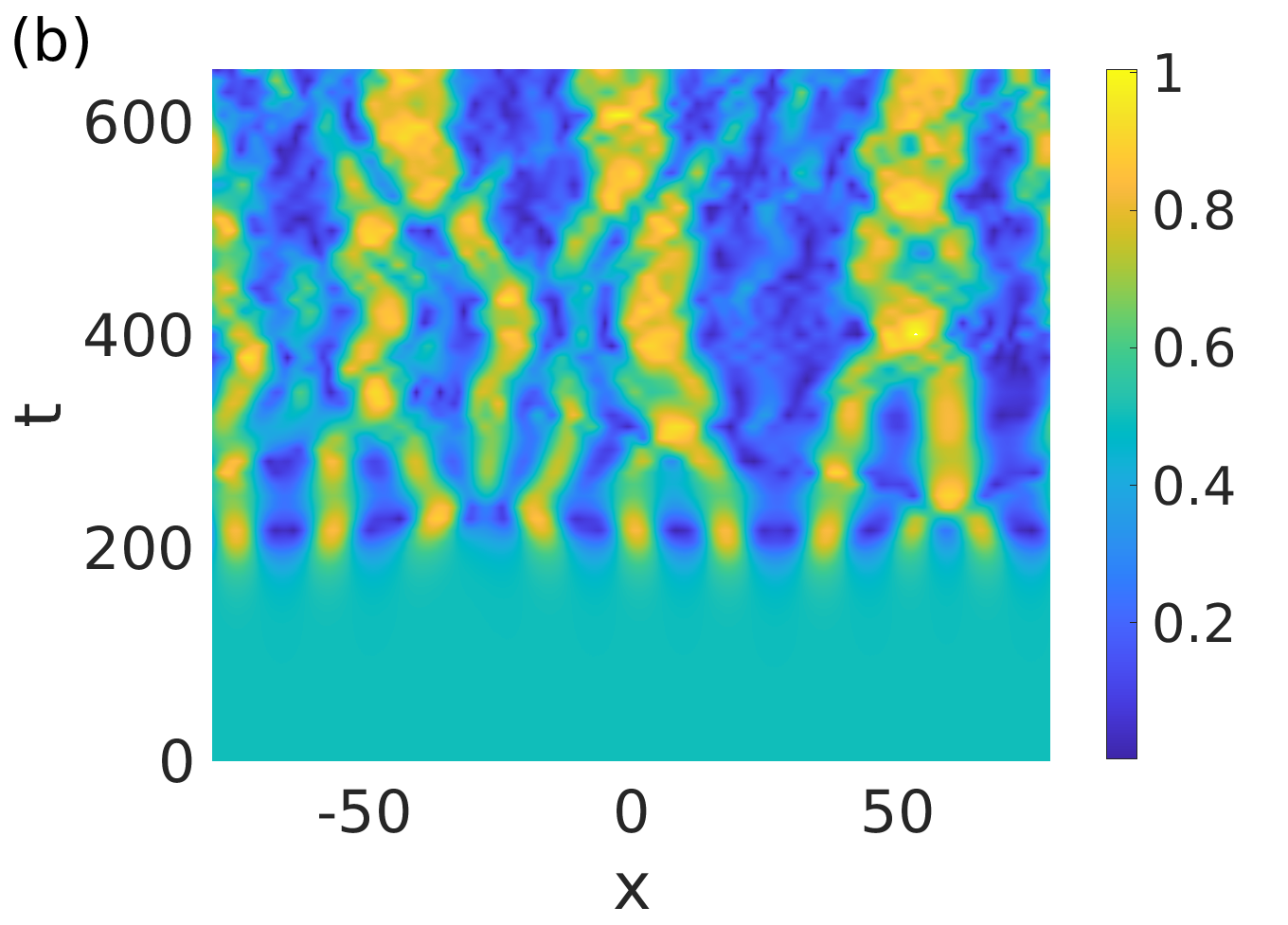} }
\caption{(a) The growth rate in ($\gamma, \delta$)-plane for $A=0.5$. Colorbar corresponds to the maximal values of
the growth rate. (b) The typical dynamics of the slightly perturbed plane waves for ($\gamma, \delta, A$)=(1,1,0.5).
}
\label{fig-5ab}
\end{figure}

In Figure 5(b), the dynamics of slightly perturbed plane waves are shown for specific values of $\gamma$, $\delta$, and $A$, presumably ($\gamma, \delta, A$)=(1,1,0.5). This figure likely captures the evolution of the system in the nonlinear phase of MI, where droplets are formed, interact, and eventually merge. The visualization offers a dynamic perspective on how MI leads to the generation of distinct, localized structures within the condensate, transitioning from initial instabilities to complex, interacting wave patterns. This behaviour reflects the critical role of MI in driving the formation and dynamics of droplet states in quasi-1D BECs.

\begin{figure}[htbp]
\centerline{ \includegraphics[width=5cm]{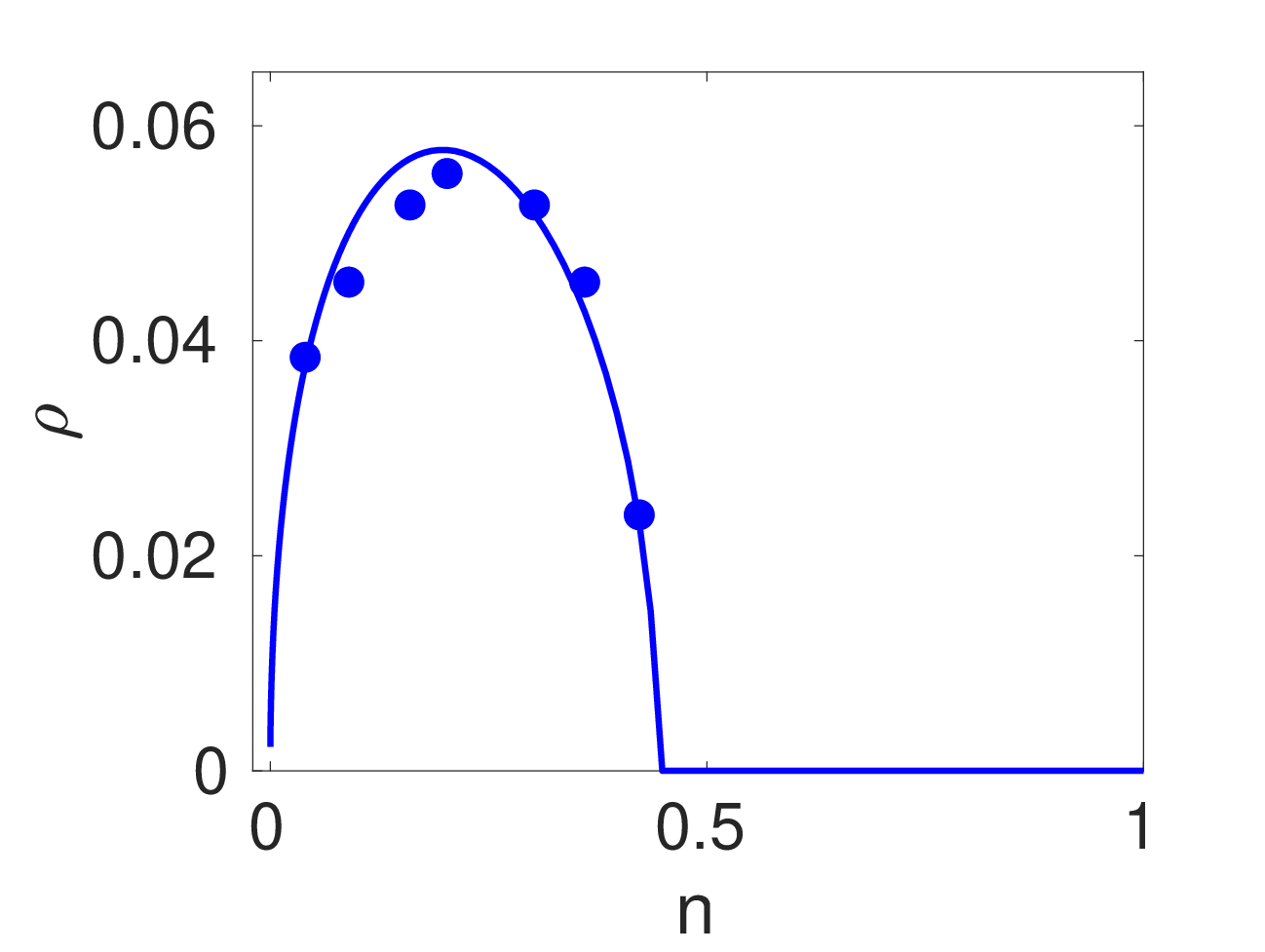} }
\caption{The number of generated QDs per unit length $\rho$ (solid line), found from the MI theory, and numerically simulated values (circles) for ($\gamma, \delta$)=(1,1). 
}
\label{fig-6}
\end{figure}

Instability causes the uniform state to evolve into a sequence of QDs. Figure~\ref{fig-6} displays a comparison of the quantum droplets generated per unit length across varying density values $n=A^2$. The distance between these QDs, or the wavelength of the most rapidly expanding modulation, is denoted by  $d=2 \pi / k_{\mathrm{max}}$. Consequently, the number of QDs per unit length is defined as $\rho = N_{\mathrm{QDs}} / L$, with $N_{\mathrm{QDs}}= L/d$ representing the total count of QDs formed. Here, $L$, the size of the simulation domain, is significantly larger than $1 / k_{\mathrm{max}}$ \cite{Mithun2020, Otajonov2022}.

	To experimentally generate a chain of droplets, we propose the following procedure. Initially, a stable, uniform BEC distribution with high density is established. Subsequently, a rapid reduction in density $n$ is achieved by loosening the transverse trapping constraints, see Fig.~\ref{fig-3ab}(b). This action facilitates the formation of a droplet chain, depicted in Fig.~\ref{fig-5ab}(b). In the context of a quasi-1D system, an alternative approach involves implementing an interaction quench via the Feshbach resonance technique. A sudden decrease in $\delta$ can transition the system from a stable to an unstable state, thereby inducing the formation of a droplet chain through modulational instability, see in Fig.~\ref{fig-4ab}(a).

\section{Conclusions}
\label{sec:conc} 

	In conclusion, our study thoroughly explored MI in Bose-Einstein condensates modelled by a modified quasi-1D GPE, emphasizing the role of mean-field effects and quantum fluctuations. We have identified the parameters that depend on MI's emergence by applying linear stability analysis to slightly modulated plane wave solutions. 

We also obtained the dependence of the number of generated QDs on different values of density through numerical simulations. In further evaluations of the plane wave dynamics, we observed the significant interactions and merging, indicating a route to larger, stable condensate structures. Our findings reveal a close relationship between GPE parameters, MI initiation, and droplet development, highlighting the balance necessary for condensate stability and BEC control.

\section*{Acknowledgements}
This work has been funded from the state budget of the Republic of Uzbekistan.


\end{document}